\documentclass[dvips,11pt]{article}

\usepackage{graphicx,times,latexsym,amssymb,overcite}

\begin{document}
\DeclareGraphicsExtensions{.pdf,.jpg,.png}

\title{
Tomography and spectroscopy as quantum computations
}

\author{
C\'esar Miquel$^1$, Juan Pablo Paz$^1$, Marcos Saraceno$^2$,\\
Emmanuel Knill$^3$, Raymond Laflamme$^4$ and Camille Negrevergne$^3$\\
\normalsize \emph{$^1$ Departamento de F\'{\i}sica,
FCEN, UBA, Pabell\'on 1, Ciudad Universitaria, 1428 Buenos Aires, Argentina}\\
\normalsize \emph{$^2$ Unidad de Actividad F\'{\i}sica, Tandar, CNEA
Buenos Aires, Argentina}\\
\normalsize \emph{$^3$ Los Alamos National Laboratory, 
MS B288, Los Alamos, NM 87545}\\
\normalsize \emph{$^4$ Los Alamos National Laboratory, 
MS B265, Los Alamos, NM 87545
}
}
\date{}

\maketitle



%
%

\def\bra#1{\langle#1|}
\def\ket#1{|#1\rangle}
\makeatletter
\def\lsim{\compoundrel<\over\sim}
\def\compoundrel#1\over#2{\mathpalette\compoundreL{{#1}\over{#2}}}
\def\compoundreL#1#2{\compoundREL#1#2}
\def\compoundREL#1#2\over#3{\mathrel
  {\vcenter{\hbox{$\m@th\buildrel{#1#2}\over{#1#3}$}}}}
\makeatother

%
%




%
%


{\bf 
Determining the state of a system and measuring properties of its
evolution are two of the most important tasks a physicist faces. For
the first purpose one can use tomography \cite{tomography}, a method
that after subjecting the system to a number of experiments determines
all independent elements of the density matrix.  For the second task,
one can resort to spectroscopy, a set of techniques used to determine
the spectrum of eigenvalues of the evolution operator.  In this
letter, we show that tomography and spectroscopy can be naturally
interpreted as dual forms of quantum computation. We show 
how to adapt the simplest
case of the well-known phase estimation quantum algorithm to perform 
both tasks, giving it a natural interpretation as a
simulated scattering experiment.  We show how this algorithm can be
used to implement an interesting form of tomography by performing a
direct measurement of the Wigner function (a phase space distribution)
of a quantum system. We present results of such measurements performed
on a system of three qubits using liquid state NMR quantum computation
techniques in a sample of trichloroethylene.  Remarkable analogies
with other experiments are discussed.}

The basic idea discussed in this letter can be described in terms of
the quantum algorithm represented by the circuit shown in Figure 1
(see [\citen{QCgeneral}] for an introduction to quantum circuits and
algorithms): A system, initially in the state $ \rho$, is brought in
contact with an ancillary qubit prepared in the state
$|0\rangle$. This ancilla acts as a ``probe particle'' in a scattering
experiment. The algorithm is: i) Apply an Hadamard transform to the
ancillary qubit (where $H|0\rangle=(|0\rangle+|1\rangle)/\sqrt 2$,
$H|1\rangle=(|0\rangle-|1\rangle)/\sqrt 2$), ii) Apply a
``controlled--$ U$'' operator (does nothing if the state of the
ancilla is $|0\rangle$ but applies $ U$ to the system if the ancilla
is in state $|1\rangle$), iii) Apply another Hadamard gate to the
ancilla and perform a {\it weak} measurement on this qubit detecting
its polarization (i.e., measuring the expectation values of Pauli
operators $\sigma_z$ and $\sigma_x$). The above circuit has the
following remarkable property:
\begin{equation}
\langle\sigma_z\rangle=\mbox{Re}[ \, \mbox{Tr}(  U \rho) \, ], \
\langle\sigma_x\rangle=-\mbox{Im}[ \, \mbox{Tr}(  U \rho) \, ].
\end{equation}
Thus, the final polarization measurement reveals a property
determined both by the initial state $ \rho$ and the
unitary operator $  U$. 
Versions of the above idea play an important role in many quantum
algorithms.  For pure input states, it occurs in Kitaev's solution to
the Abelian stabilizer problem (a generalization of the factoring
problem)~\cite{kitaev:qc1997a}.  This was adapted by Cleve
\emph{et al.}~\cite{cleve:qc1997b} to revisit most quantum algorithms,
with the circuit of Figure 1 being the simplest instance (see 
[\citen{Lloyd99}] for yet another presentation of the algorithm as a 
tool for physics simulations).  The
extension to mixed states and a version of Equation 1 is
in [\citen{knill:qc1998a}]. 

Here, we observe that as a result of Equation 1, this circuit can be
used for dual purposes: We can use it to extract information on the
operator $ U$ if we know the state $ \rho$. Alternatively, we can use
it to learn about the state $ \rho$ by using some specific operators
for $ U$. In this sense, this circuit can be adapted either as a
tomographer or as a spectrometer which, therefore, are dual faces of
this quantum computation.  Moreover, it is interesting to think of the
above algorithm as simulating a scattering experiment: The ancillary
qubit plays the role of a probe particle: it interacts with the
scatterer (the system, initially in state $\rho$) and is later
detected. From the statistics gathered by the detector measuring the
polarization of the probe, we can either reconstruct the state of the
scatterer (if we know about the interaction) or learn about the
interaction (if we know about the state of the scatterer). Describing
the algorithm in physics language helps in establishing analogies
between the above proposal and some already performed experiments
that, interestingly enough, can be interpreted as special instances of
the algorithm (see below).

Let us first discuss how to use this to build a spectrometer.  The
idea is simple: For example, preparing the system in a completely
mixed initial state with $ \rho= I/N$ ($N$ is the dimensionality of
the space of states of the system), the final measurement is $\langle
\sigma_z\rangle=\mbox{Re}[\, \mbox{Tr}( U)\, ]/N$.  This is directly
proportional to the trace of $U$, the sum of all its eigenvalues, and
therefore has information about the spectrum. Moreover, we can build a
real spectrometer by simply adding an extra register (with $n_1$
qubits) and a controlled Fourier transform. With this new circuit,
shown in Figure 2, the final polarization is
$\langle\sigma_z\rangle=g(2E)=\mbox{Re}[\,\sum_t^T \exp(4\pi i Et/T)
\,\mbox{Tr}( U^t)\,]/NT$ (where $T=2^{n_1}-1$). The function $g(E)$ is
the Fourier transform of $\mbox{Tr}(U^t)$, which is nothing but the
spectral density of $U$ (smoothed over a scale $2\pi/T$).  One can use
this same idea to design circuits to measure, for example, the
structure function, defined as the Fourier transform of $|{\mbox
Tr}U^t|^2$, characterizing the spectral correlations and the level
spacing statistics of $U$, which are of interest in studies of quantum
chaos \cite{CircuitsUS} (some other related constructions and
efficient networks can be found in [\citen{Gerardo}]).  It is also
important to notice that the above is a rather general spectrometer
since it can be adapted to study properties of non--Unitary
operators. For example, by adding an extra ancillary register one can
use it as a spectrometer for any operator obtained as a weighted sum
of unitary operators.  We emphasize that the above is a
\emph{simulation} of a spectrometer where one has complete control
over the operator $U$ that acts on the system.  This is the case for
any physics simulation, either quantum or classical, where our
simulation is intended to be used for the purpose of exploring models.
On classical computers, the task of providing spectral information
as accomplished by this quantum algorithm appears to be exponentially hard
in general.  Of course, complete determination of the spectrum of $U$
is still inefficient, unless the eigenvalues are hugely degenerate.
But using our simulated spectrometer one can efficiently determine
important specific properties of the spectrum, such as directly
sampling the spectral density or the structure function.

The circuit in Figure 1 can be adapted for another important 
purpose: state tomography. Every time we run the algorithm 
for a known operator $  U$, we extract information about 
the state $ \rho$. Doing so for a complete basis of 
operators $\{  A(\alpha), \alpha=1,\ldots,N^2-1\}$ one gets 
complete information and determines 
the full density matrix. Different tomographic schemes are 
characterized by the basis of operators $  A(\alpha)$ one uses. 
In NMR spectroscopy, for example, a convenient 
basis set is formed by all tensor products
of Pauli operators (the basis of the product operator 
expansion \cite{sorensen:qc1983a}). This kind of tomography is
particularly well suited in this context since all 
$  A(\alpha)$ are easily reached from observables by a sequence 
of r.f. pulses and periods of free evolution \cite{cory:qc2000a}.
However, as the above is a generic tomographer one 
can explore other choices. In particular, we can  
use it to implement another important tomographic scheme 
characterized by the operators  
\begin{equation}
  A(q,p)={1\over 2N} \tilde U^q   R \tilde V^{-p}
\exp(i2\pi pq/2N),\label{aqpdiscrete}
\end{equation}
where $\tilde U$ is the operator producing a cyclic
shift in the computational basis ($\tilde U|q\rangle=|q+1\rangle$),
$\tilde V$ is the shift in the basis related to the
computational one via the discrete Fourier transform 
and $  R$ is the reflection operator 
($ R|q\rangle =|N-q\rangle$). By doing this, we measure the discrete
Wigner function of the state of the system, which is defined as
$W(\alpha)=\mbox{Tr}( A(\alpha) \rho)$ (we use $\alpha=(q,p)$).  This
function is the basic tool to represent the state of a quantum system
in phase space, the natural arena of classical physics \cite{Wigner}
and has often been used to understand the nature of the quantum to
classical transition \cite{juanpablo}. Its use is common in various
areas of physics, but has been mostly restricted to continuous systems
(discrete Wigner functions have been used only recently in the context
of quantum computation \cite{WignerUS}).  The defining properties of
the Wigner function are: (P1) $W(\alpha)$ is real valued, (P2) Inner
products between states $\rho_1$ and $\rho_2$ can be obtained as
$\mbox{Tr}(\rho_1\rho_2)=N\sum_{\alpha} W_1(\alpha)W_2(\alpha)$, (P3)
Adding the value of $W(\alpha)$ over all points in any phase space
line gives a positive number which is the probability of measuring a
physical observable. These are consequences of interesting properties
of the phase space point operators $ A(\alpha)$
\cite{WignerUS,Hannay}.  One of these properties is the completeness
of the set $ A(\alpha)$ (a complete orthogonal set is obtained with
$\alpha$ restricted to the first $N\times N$ sub-grid of the phase
space but it is necessary to define $W(\alpha)$ in a grid of $2N\times
2N$ points for $P3$ to hold \cite{WignerUS,Hannay}).  One can expand
the state $ \rho$ in terms of this set as $\rho=N\sum_{\alpha}
W(\alpha) A(\alpha)$.  The coefficients of the expansion are the
Wigner function of $\rho$.

Every time one runs this algorithm with $U=  A(\alpha)$ 
a direct measurement of $W(\alpha)$ is obtained (for 
any given $\alpha$). Before presenting results of such 
measurement for a simple system, 
it is worth mentioning that the same scheme can be used to measure 
$W(\alpha)$ in more general cases. 
For a continuous system $W(\alpha)$ is  
defined as the expectation value of 
$A(\alpha)=D^\dagger(\alpha)  R D(\alpha)/\pi\hbar$ where 
$D(\alpha)$ is the phase space displacement operator 
$D(\alpha)=\exp(i(p\hat Q-q\hat P)/\hbar)$.
Therefore, to measure $W(\alpha)$ we need 
to run the algorithm with the system in the state
$D(\alpha) \rho D^\dagger(\alpha)$ (obtained simply 
by displacing $\rho$) and use $U = R$ \cite{CQEDnote}. 
Measuring directly the Wigner function has been
the goal of a series of experiments 
in various areas of physics (all dealing with continuous systems
\cite{WignerMeas}). It is remarkable that our measurement scheme   
describes the recent experiment that 
determined $W(\alpha)$ for the electromagnetic 
field in a cavity QED setup \cite{DavidovichLut,Harocheetal}. 
In this case the system is the mode of the field stored in a high--$Q$
cavity and the ancillary qubit is a two level atom. The measurement
of the Wigner function is done via a scattering experiment that  
directly follows the steps described in Figure 1:
i) The atom goes through a Ramsey zone that has the effect of
implementing an Hadamard transform. An r.f. source is
connected to the cavity displacing the field (by an
amount parametrized by $\alpha$) and preparing the state $D(\alpha)
\rho D^\dagger(\alpha)$;
ii) The atom goes through the cavity interacting dispersively with
the field in such a way that only 
if the atom is in state $|e\rangle$ does it acquire a phase shift
of $\pi$ per each photon in the cavity (i.e., this interaction
is a controlled--$\exp(-i\pi\hat N)$ gate, where $\hat N$
is the photon number, which is nothing but a controlled reflection).
iii) The atom leaves the cavity entering a new Ramsey zone and
is finally detected in a counter
either in the $|g\rangle$ or $|e\rangle$ state.
The Wigner function is measured as the difference between 
both probabilities: $W(q,p)= 2 (P(e)-P(g))/\hbar$. 
As we see, this cavity--QED experiment is a concrete realization
of the general tomographic scheme described above. 

For a system with a finite dimensional Hilbert space  
the algorithm 
can be efficiently decomposed as a sequence of elementary steps:
All controlled-$\tilde U$, $\tilde V$ and $  R$ operations
can be implemented via efficient networks like the ones shown
in [\citen{networks}] which require a Poly$(\log(N))$ 
number of elementary gates. For small $N$ these networks
are very simple. We implemented the measurement of 
$W(\alpha)$ for $N=4$ (two qubits) for a variety
of initial states. In this case $  R$ is a controlled
not (CNOT) gate (where the control is in the least significant qubit).
$\tilde U$ is the same CNOT followed by a bit flip in the
control. Analogously, $\tilde V$ is a sequence
of controlled phase gates. The complete circuit
has at most one Toffoli gate and several two qubit gates. Figure 3
shows the results of the measurement of the Wigner function for 
all four computational states of a
two qubit system. Ideally, $W(\alpha)$ for the state 
$|q_0\rangle$ is nonzero only on the vertical strip at $q=2q_0$,
where it is equal to $1/2N$ and on the strip at $q=2q_0\pm N$
where it oscillates as $(-1)^p/2N$. These 
oscillations correspond to interference between
the state and its mirror image created by the periodic boundary
conditions \cite{WignerUS}.

To measure $W(\alpha)$ we used a liquid sample of trichloroethylene 
dissolved in chloroform. This molecule has been used in several three qubit
experiments where the proton ($^1$H) and two strongly coupled  $^{13}$C
nuclei ($C_1$ and $C_2$) store the three qubits \cite{nmrexperiments}. 
We used $C_1$ as our probe particle and the
pair $H$-$C_2$ to store the state whose Wigner function we 
measure. The coupling constants
are $J_{HC_1}=200.76 \ \mbox{Hz}$, $J_{HC_2}=9.12 \ \mbox{Hz}$
and $J_{C_1C_2}=103.06 \ \mbox{Hz}$ while the $C_1$-$C_2$ chemical shift is
$\delta_{C_1C_2}=908.88 \ \mbox{Hz}$. We determined the value of 
$W(\alpha)$ for each of the independent
$16$ phase space points. Each of these circuits corresponds to a 
different sequence of r.f. pulses and delays (the number of 
pulses in each sequence depends on $\alpha$ varying 
between $5$--$17$ and taking at most $100$ms to execute, 
$\ll T_1,T_2$ of our sample).  We used temporal averaging
\cite{temporalav} to obtain, from the part of $  \rho$ that
deviates from the identity, the four pseudo-pure 
initial states whose Wigner functions are shown in Figure \ref{figure2}.
The experiments were done at room temperature 
on a standard $500$Mhz NMR spectrometer
(Bruker AM-500 at LANAIS in Buenos Aires and DRX-500 at Los Alamos).
We used a $5$mm probe tuned to $^{13}$C and $^1$H frequencies of $125.77$ MHz
and $500.13$ Mhz. 
The most important sources of errors 
come from the strong
coupling 
and the numerical uncertainty in
integrating the spectra. These results illustrate the tomographic 
measurement of a discrete Wigner function and agree very well with 
theoretical expectation.

In summary, we presented a general algorithm enabling 
us to view spectroscopy and tomography as dual tasks. 
From our algorithm follows
the construction of circuits characterizing spectral properties 
of general operators and the design of a general tomographer with which we 
can directly measure the Wigner function of the state 
of a quantum system.
The measurement strategy is an example of a 
very general tomographic scheme in terms of which one can 
interpret previous experiments performed to measure Wigner 
functions. The analogy between this
quantum computation and a scattering experiment
(that has the dual use of providing information either about the
state of the scatterer or about the interaction Hamiltonian) is, we
believe, particularly illuminating and useful.

%
%


\begin{thebibliography}{99}
\bibitem{CQEDnote} To measure the continuous Wigner function 
one can choose between two equivalent options: i) applying 
the circuit of Figure 1 with $U=D(\alpha) R D^\dagger(\alpha)$,  
or ii) apply first the operator $D(\alpha)$ displacing the 
initial state $\rho$ and later use the circuit with $U=R$. 
This later option is more convenient since in this way the 
only operation that needs to be controlled by the ancillary 
qbit is the reflection $R$ (this is precisely what is done 
in the cavity QED experiment described in the text). This 
strategy can always be applied for operators of the form
$U = G B G^\dagger$ (in such case, one can first unconditionally 
apply the operator $G$ to the initial state and later use 
a simpler circuit with $U=B$). It is interesting to notice that
this strategy can be applied to the measurement of the 
discrete Wigner function only if both $q$ and $p$ are even (some
of these peculiar features of discrete Wigner functions are 
discussed in [\citen{WignerUS}]).
\end{thebibliography}
%

%
%
\pagebreak

%
%

\begin{figure}
\centering \leavevmode
\vspace {0.5cm}
\caption{Circuit for measuring $\mbox{Re}[ \mbox{Tr}( \rho U) ]$, 
for any unitary operator $U$.}\hfill
\label{figure1}
\end{figure}

%
%

\begin{figure}
\centering \leavevmode
\vspace {0.5cm}
\caption{Circuit for evaluating the spectral density of an operator
$U$. The second register, formed by $n_1$ qubits, is prepared in the
initial state $|E\rangle$ and is subject to a controlled Fourier transform.
The second controlled operator acts on the second and third registers
mapping the computational states $|t\rangle_2\otimes |n\rangle_3$
into $|t\rangle_2\otimes U^t|n\rangle_3$.}
\label{figurespec}
\end{figure}

%
%

\begin{figure}[h]
\centering \leavevmode
\caption{Measured Wigner functions for the four computational states
of a two--qubit system (built with a liquid sample of TCE in an NMR 
spectrometer).
Ideally, these
Wigner functions should be nonzero only on two vertical strips where they
take values which are $\pm 1/8$. Experimental results show small deviations
from these values (with a maximum error of $15\%$.}
\label{figure2}
\end{figure}

\vfill

%
%

\pagebreak
\thispagestyle{empty}
\parbox[t]{17cm}{Figure 1. Juan Pablo Paz {\tt <paz@df.uba.ar>}}
\vspace{4cm}

\begin{figure}[h]
\centering \leavevmode
\includegraphics[width=17cm]{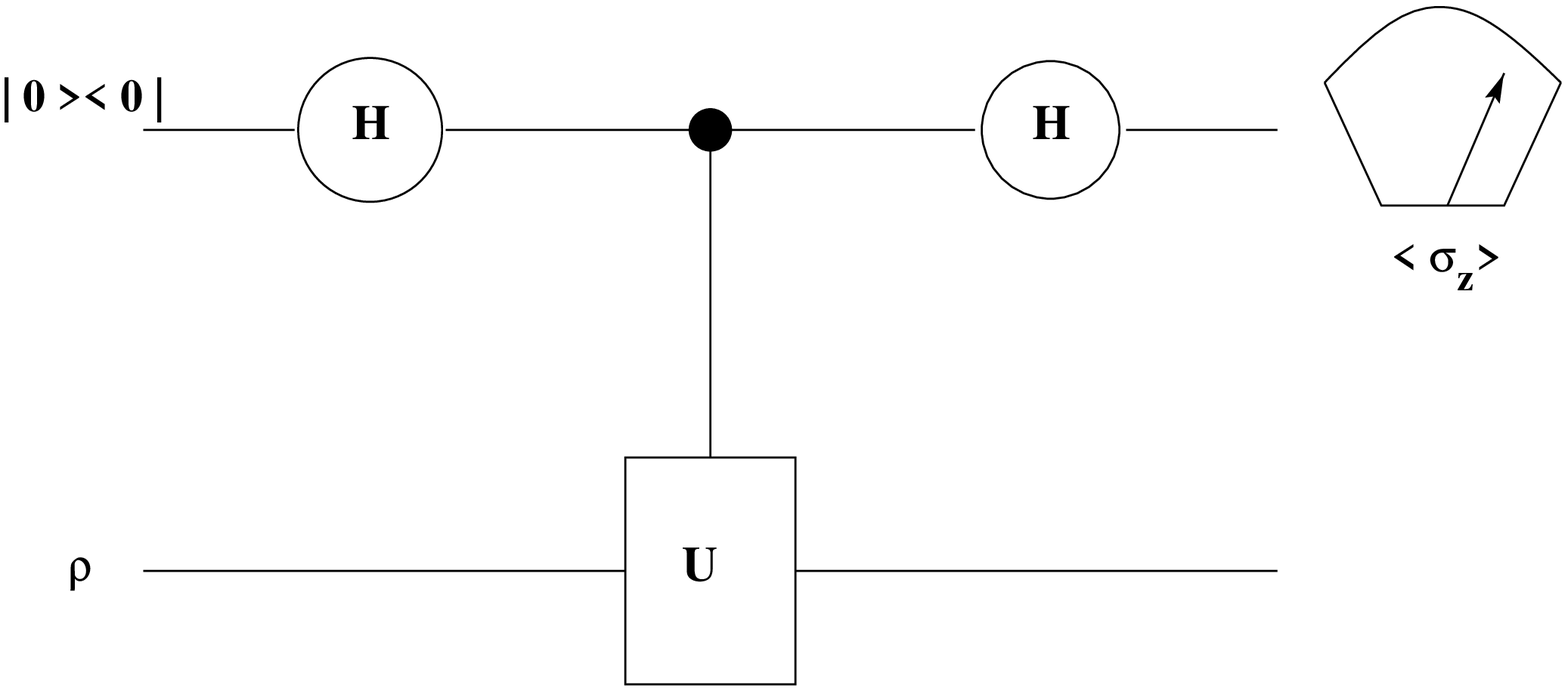}
\end{figure}

\cleardoublepage
\pagebreak

\thispagestyle{empty}
\parbox[t]{17cm}{Figure 2. Juan Pablo Paz {\tt <paz@df.uba.ar>}}
\vspace{3cm}

\begin{figure}[h]
\centering \leavevmode
\includegraphics[width=17cm]{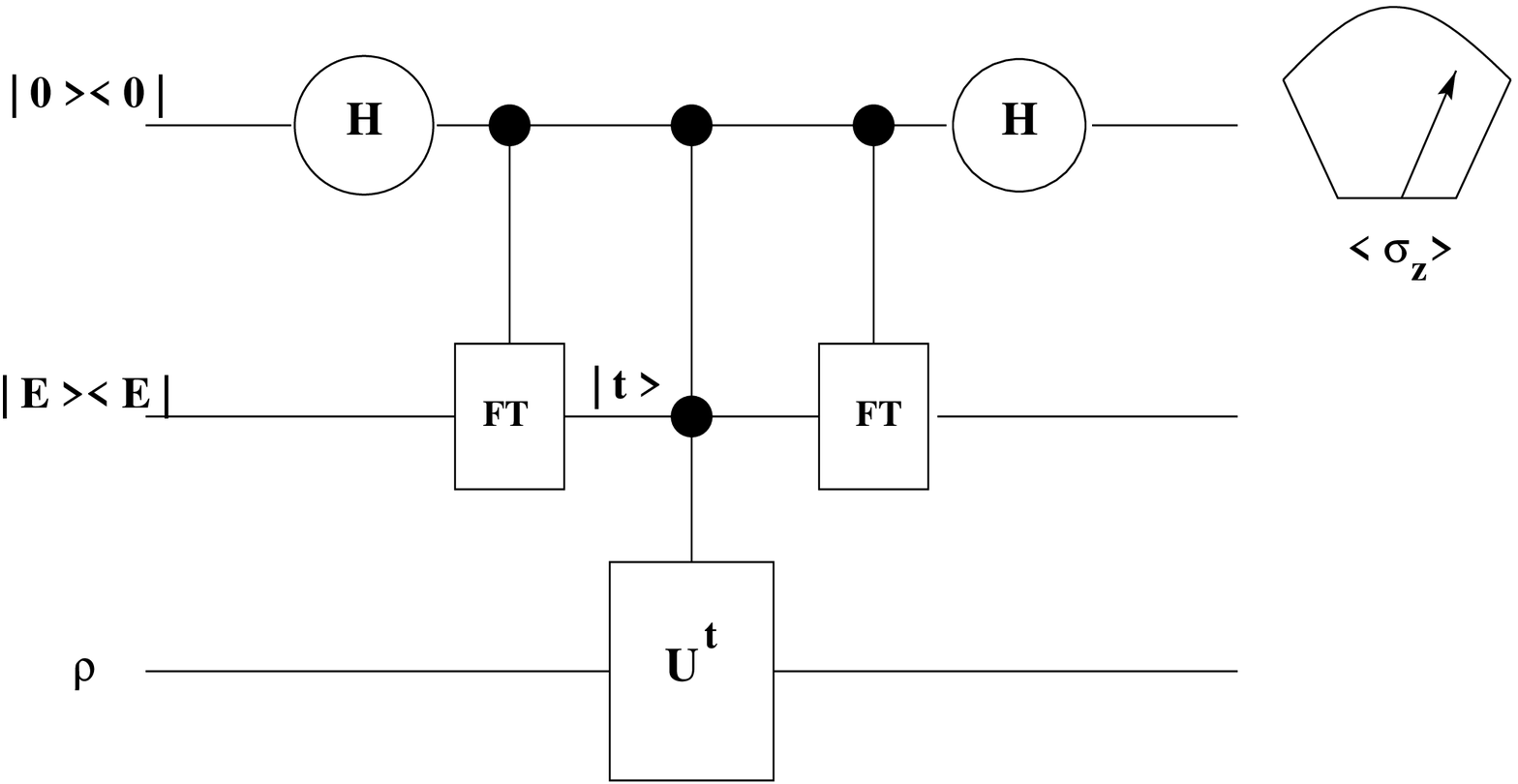}
\end{figure}

\pagebreak

\thispagestyle{empty}
\parbox[t]{17cm}{Figure 3. Juan Pablo Paz {\tt <paz@df.uba.ar>}}
\vspace{2cm}
\begin{figure}[h]
\centering \leavevmode
\includegraphics[width=13cm]{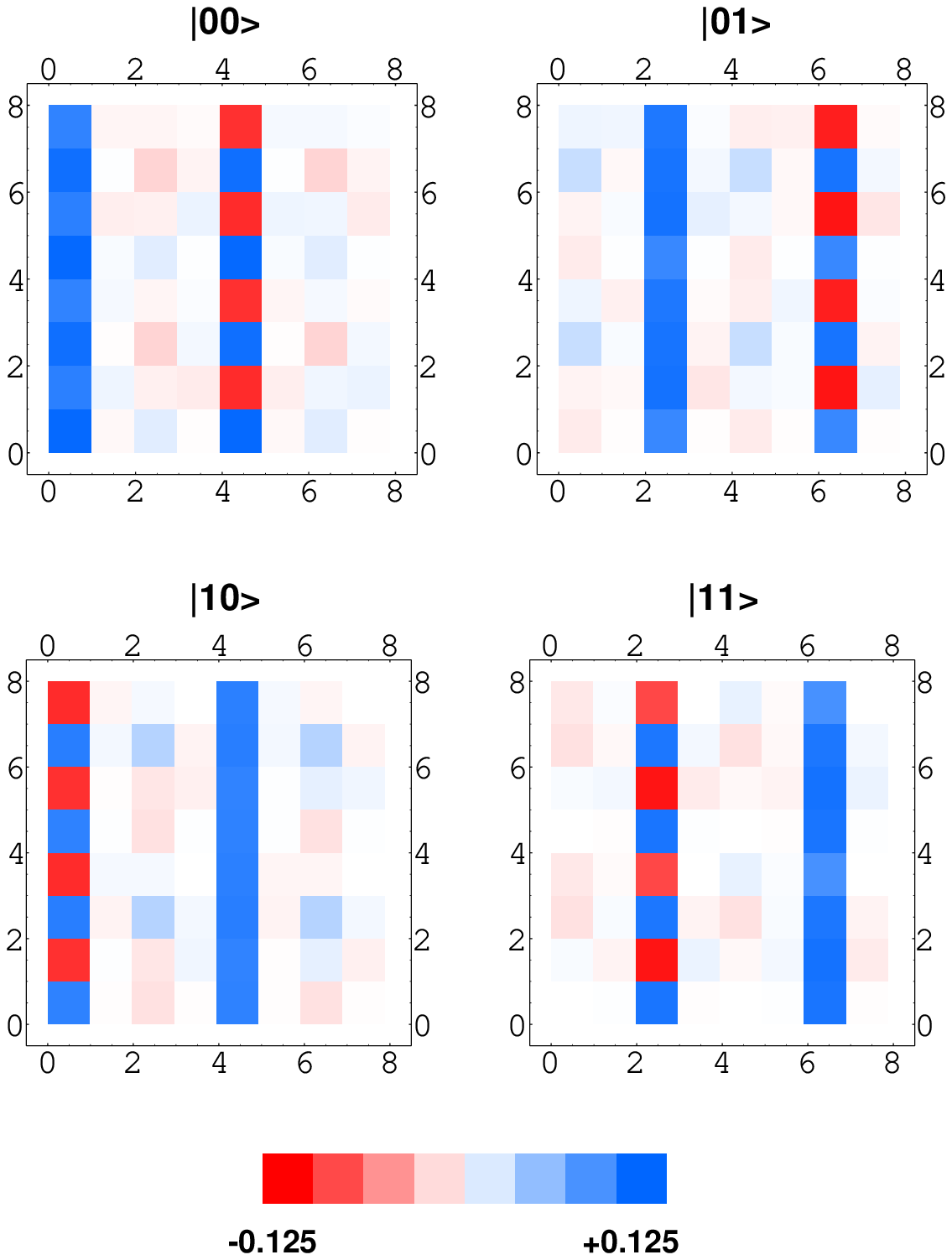}
\end{figure}

\end{document}